# THE GAMMA TRANSITIONS, LEVEL DENSITIES AND GAMMA STRENGTH FUNCTION OF ISOTOPES $^{172}$Yb AND $^{153}$Sm

Vuong Huu Tan[1], Pham Dinh Khang[2], Nguyen Xuan Hai[3], Ho Huu Thang[3], Dao Manh Trinh[4], Nguyen An Son[4]

[1] *Vietnam Agency for Radiation and Nuclear Safety, 113 Tran Duy Hung, Hanoi*
[2] *Nuclear Training Center, 140 Nguyen Tuan, Hanoi*
[3] *Nuclear Research Institute, 01 Nguyen Tu Luc, Dalat*
[4] *University of Dalat, 01 Phu Dong Thien Vuong, Dalat*

**Abstract:** This report is presented the level densities and gamma strength function which is extracted from primary gamma transitions of (n,2γ) measurement on the Dalat Nuclear Research Reactor (DNRR). The comparisons with others results and model calculations, there are outside the scope of uncertainty. These research results are showed that having of complement research for deformed nucleus in order to evaluate nuclear structure models.

***Keywords***: *level density, primary gamma transitions, (n,2γ) reaction.*

## I. INTRODUCTION

The level density and radiative strength function are fundamental nuclear properties and important input parameters in large network calculations of the nucleon synthesis of heavy elements. The main experimental information about the γ-strength function is obtained from photo absorption experiments where reaction as (γ,xn) and (γ,xp) are investigated. It has been found that in the energy region from ~8 to 20 MeV [6, 7]. The region below the particle separation threshold (<$B_n$) where the particle emission is forbidden. The main experimental information about strength function in this case is provided by gamma transitions emitted from compound nuclei formed in different reactions. Due to the absence of coulomb barrier for neutrons, the most of the data have been obtained from thermal and resonance neutron capture reactions because of high probability that reaction proceeds through a compound nucleus at these energies. The most comprehensive systematic of gamma strength functions [9], based on neutron capture reactions is presented in the RIPL-2 data base. It gives the absolute numbers of gamma strength functions close to the neutron binding energy (~$B_n$-1 MeV) as well as recommended energy dependence below the particle separation threshold [6, 7, 9]. However, the accuracy of this systematic is estimated to be a factor of three times. Strength functions of magnetic dipole and electric quadrupole gamma transitions are even more uncertain because of the lack of the experimental information.

There are two main problems with obtaining gamma strength functions from emission gamma spectra of nuclear reactions.

*1. Absolute values of the strength function are obtained from absolute intensities of primary gamma transitions from resonance neutron capture reactions. However, due to the lack of experimental information, the absolute intensities suffer from high uncertainties.*

*2. The shape of the emission gamma spectrum depends not on the gamma strength function only but also on the level density of the nucleus populated by these gamma transitions. Therefore the knowledge of the level density is crucial for the study of the gamma strength function.*

Therefore, we look for some other experimental approaches to be able to increase the accuracy of the experimental gamma strength function below the particle separation threshold [8].

In this work, we study experimental level density and gamma strength function of $^{153}$Sm and $^{172}$Yb nucleus by gamma two step cascade method at Dalat Nuclear Research Reactor (DNRR) [1, 2].

## II. EXPERIMENT

The experiments were performed at DNRR with thermal neutron flux at sample position is about $10^6$ n.cm$^2$.s$^{-1}$, the rich isotope targets are YbO$_2$ and SmO$_2$, the gamma from $^{171}$Yb(n,2γ)$^{172}$Yb and $^{152}$Sm(n,2γ)$^{153}$Sm reactions had been measured by coincidence spectrometer which is based on event-event method. The data was analyzed according to $E_1+E_2 \leq B_n-E_f$ conditions [8]. $E_1$ and $E_2$ are energies of pairs of gamma cascade, $B_n$ is neutron binding energy, and $E_f$ is energy of final state of cascades. The 65 primary gamma transitions of $^{172}$Yb and 214 primary gamma transitions were founded.

The primary gamma strength function $f = \Gamma_{\lambda i}/(E_\gamma^3 \times A^{2/3} \times D_\lambda)$ and level density determine the total radiative width of the compound state and cascade intensity I, obtained in the following way [8]:

$$\Gamma_\lambda = \langle \Gamma_{\lambda i} \rangle \times m_{\lambda i} \quad (1)$$

$$I_{\gamma\gamma} = \sum_{\lambda,f} \sum_i \frac{\Gamma_{\lambda i}}{\Gamma_\lambda} \frac{\Gamma_{\lambda f}}{\Gamma i} = \sum_{\lambda,f} \frac{\Gamma_{\lambda i}}{\langle \Gamma_{\lambda i} \rangle m_{\lambda i}} n_{\lambda i} \frac{\Gamma_{if}}{\langle \Gamma_{if} \rangle m_{if}} \quad (2)$$

Here $_i$, is the partial width of -transition with energy E, A is the nucleus mass, and D is the spacing between decaying levels. The values of the total and partial gamma-widths are set for the compound state and the cascade's intermediate level i, respectively; m is the total number of the excited levels, and n is the number of excited levels below given level or i in the energy interval of the average cascade intensity determination.

The level density is defined as a function of excited energy as equation (4).

$$\rho(U) = \frac{d}{dU} N(U) \quad (3)$$

where $N(U)$ is number of excited levels in the range energy ≤U. The simplest model which has been developed a long time ago is based on the Fermi-gas approximation which assumes that a nucleus consists of non-interacting Fermi-particles freely moving in the sphere of the nuclear volume. The gas is degenerate, i.e. all particles populate only lowest levels consistent with Pauli principle.

In framework, the back-shifted Fermi gas model was used, the excitation energy U and level density $\rho(U,J)$ of a nucleus with a given angular momentum J can be expressed as [3, 9]:

$$\rho(U,J) = \frac{2J+1}{24\sqrt{2}\sigma^3 a^{1/4}(U-\Delta+t)^{5/4}} \exp\left[2\sqrt{a(U-\Delta)} - \frac{J(J+1)}{2\sigma^2}\right] \quad (4)$$

Where $a$ is level density parameter, $\Delta$ is the excitation energy shift, t is the nuclear temperature, and $\sigma$ is the spin cutoff parameter. That parameters are attract from reference [5], the calculate results are not suitable with experiment. The parameters have been free to equation (4) and fitted with experimental data; the result agreed in excited energy region ~0.7$B_n$.

III. RESULTS

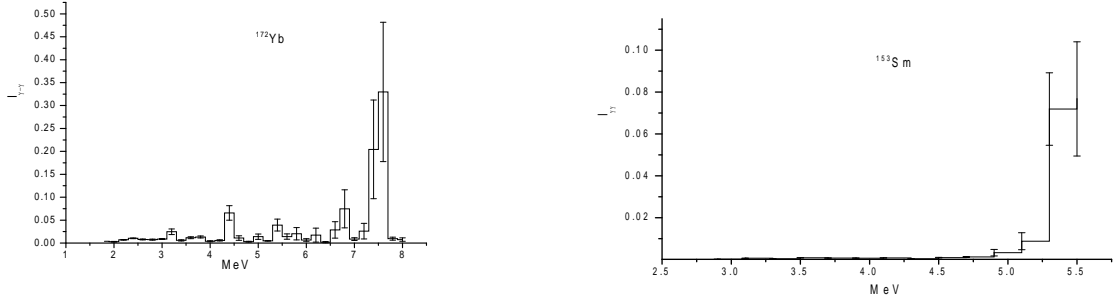

*Fig. 1. Primary gamma intensity transitions.*

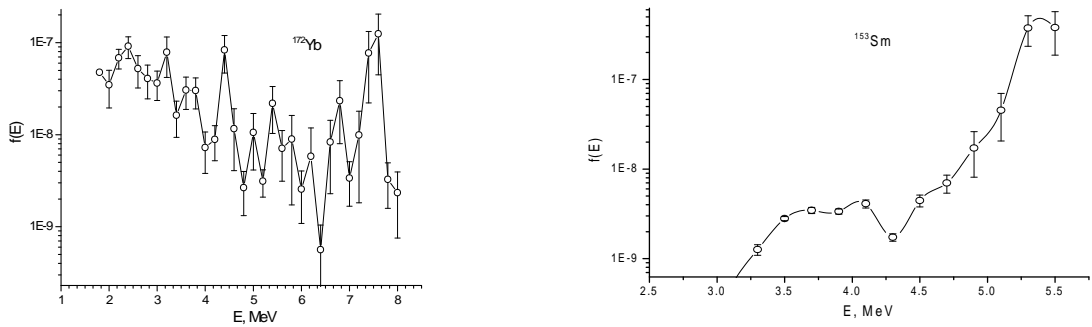

*Fig.2. Primary gamma strength functions.*

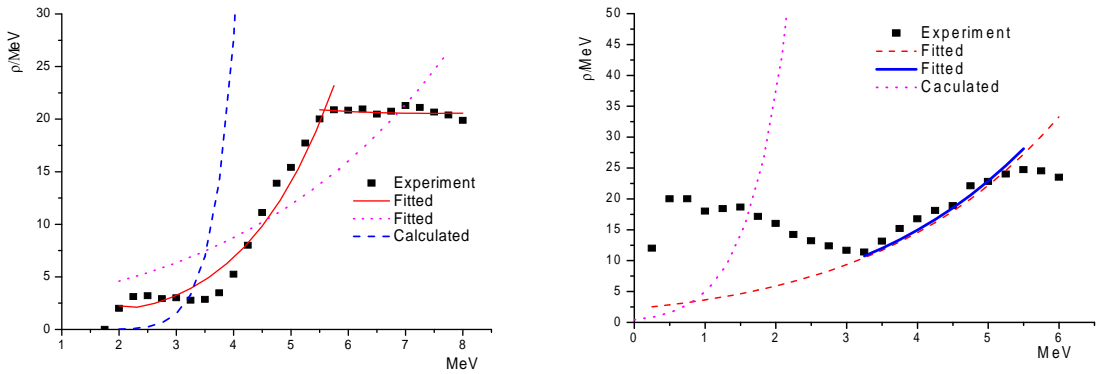

*Fig.3. Excited level density after primary gamma transitions.*

In general, about 30% measured primary gamma energy are like in the reference [4]. The gamma transitions have energy from 0.3÷6 MeV (Fig.1), but in the reference have from 3÷7 MeV [10, 11]. That might be due to from difference of experiment. In the energy region from ≥0.7$B_n$ the level density is decreased. The single particle model predicts the increasing of level density to excited energy, but nucleus are combined from nucleons, so that when excited energy are enough high, the effect collective is increased, alternative the effect single is decreased it effect to decrease level density of exciting. The limited resolution of detectors can not detect all excited level, besides the efficiency is tiny in upper 3 MeV regions.

The primary gamma transition intensity is fitted by Lorentz function like expression (5), this like that in Kopecky theorem [9] (Fig.1). The giant and Pygmy resonances have been founded for both $^{153}$Sm and $^{172}$Yb.

$$I_\gamma(\varepsilon_\gamma) = \alpha\sigma\Gamma \frac{U\Gamma}{(\varepsilon_\gamma^2 - E^2)^2 + (\Gamma\varepsilon_\gamma)^2} \text{ (MeV}^{-3}\text{)}, \tag{5}$$

Where: the $\sigma$, $E$, and $\Gamma$ are the peak cross section, energy and width of resonance, respectively; factor $\alpha$ depends on experimental data.

## IV. CONCLUSSION

The level density and gamma strength function are very old problems which have been widely studied since the early days of nuclear physics. The Fermi-gas expressions are still frequently used even if they do not account for all the physical properties, especially for collective effects. Since it is well established that these collective effects play a key role for low excitation energies and vanish with increasing excitation energy, old level density expression can only be satisfactory over a very narrow range of excitation energy. For high energy calculations required by current nuclear research programs, modern expressions have begun to be developed to include more precisely the collective effects and describe the way they vanish.